\documentclass[superscriptaddress,showkeys,preprintnumbers,amsmath,amssymb,floatfix,prd]{revtex4}
\usepackage{graphicx}% Include figure files
\usepackage{dcolumn}% Align table columns on decimal point
\usepackage{bm}% bold math
\usepackage{hyperref}
\hypersetup{colorlinks,citecolor=cyan}

\begin{document}

\preprint{}

\title{Deflection Angle of Regular Black Holes in Nonlinear Electrodynamics: Gauss-Bonnet Theorem, Time Delay, Shadow, and Greybody Bound}

\author{Susmita Sarkar}
\email{susmita.mathju@gmail.com}
\affiliation{School of Applied Science and Humanities, Haldia Institute of Technology, Haldia-721606, Purba Mednipur, West Bengal, India}

\author{Nayan Sarkar}
 \email{  nayan.mathju@gmail.com}
\affiliation {Department of Mathematics, Karimpur Pannadevi College, Karimpur-741152, Nadia, West Bengal, India}

\author{Hasrat Hussian Shah}
\email{hasrat@mail.ustc.edu.cn}
\affiliation{Department of Mathematical Sciences, Balochistan University of Information Technology Engineering and Management Sciences, Quetta-87300, Pakistan}

\author{Pankaj Balo}
\email{pankajbalo@gmail.com }
\affiliation{Department of Mathematics, Jadavpur University, Kolkata-700 032, West Bengal, India}

\author{Farook Rahaman}
\email{rahaman@iucaa.ernet.in}
\affiliation{Department of Mathematics, Jadavpur University, Kolkata-700 032, West Bengal, India}

\date{\today }

\begin{abstract}
In this article, we study the weak gravitational lensing in the background of regular, static, spherically symmetric black hole solutions of Einstein's standard general relativity coupled with nonlinear electrodynamics.  The weak deflection angles are estimated in the context of vacuum medium and plasma medium using the Gauss-Bonnet method. The obtained deflection angles decrease as the impact parameter $b$ and the charge parameter $q$ increase, while the deflection angles increase gradually with increasing values of the black hole mass $m$. Moreover, the effect of a plasma medium has increased the deflection angle than the vacuum medium scenario. We also estimate the time delay in the field of the described black holes that vanishes for $m = q =0$, i.e., the absence of the black holes. The shadow cast of the present black holes is also analyzed with respect to the impact $q$ and mass $m$, which ensures that the shadow region shrinks for increasing values of $q$ and expands for increasing values of $m$. In addition, we estimated the rigorous bounds of the greybody factor $\mathcal{T}_b$ for the described black holes and the graphical analysis ensures that the increasing charge parameter $q$ decreases the rigorous bound of  $\mathcal{T}_b$ and the increasing mass $m$ increases the rigorous bound of $\mathcal{T}_b$.

\end{abstract}

\maketitle

%%%%%%%%%%%%%%%%%%%%%%%%%%%%%%%%%%%%%%%%%%%%%%%%%%%%%%%%%%%%%%%%%%
\section{Introduction}\label{secI}
%%%%%%%%%%%%%%%%%%%%%%%%%%%%%%%%%%%%%%%%%%%%%%%%%%%%%%%%%%%%%%%%%%
Einstein's theory became the subject of numerous experimental tests since its discovery in 1915 \cite{A. Einstein}. After the groundbreaking discovery of gravitational waves and observational evidence of the event horizon with other confirmed experiments \cite{W. de Sitter, B.P, B.P 2, T. Johannsen}, it turns out that experimental results are followed by theoretical predictions of Einstein's theory of general relativity. Light bending phenomenon as it goes through a gravitational field, called as one of the predictions of general relativity \cite{A.S, L.C}. Subsequently, gravitational lenses are considered to be a very important tool in the field of astrophysics and cosmology \cite{A. Einstein 2}, it helps in investigating the structural distribution, dark matter dynamics, etc. \cite{Y. Mellier, N. Kaiser}. The angle of light deflection relies on the distance of the observer from the lens and the properties of the lens. Therefore, it indicates that one may get the information about the nature of black holes from the gravitational lensing effect of black holes. This property of the gravitational lensing leads us to distinguish among different black holes \cite{E.F, G.N, A.Y}, and may test the various theories of gravity \cite{S.G, Z.Hora, J. Badia, X.M}. 

Regular black holes are defined as a subclass of black hole solutions that admit only coordinate singularities as black hole horizons but eliminate essential singularities, ensuring finite curvature invariants throughout spacetime, particularly in the central region \cite{id92, ka01}. According to the well-known singularity theorems established by Penrose and Hawking \cite{sw73} demonstrate that, under certain conditions, the occurrence of singularities in General Relativity is unavoidable. This aligns with the fact that the earliest exact black hole solutions in General Relativity contain a singularity hidden within the event horizon. Nevertheless, it is widely believed that such singularities are nonphysical artifacts arising from classical gravitational theories and do not actually exist in nature. Indeed, quantum considerations put forward by Sakharov \cite{ad66} and Gliner \cite{eb66} indicate that spacetime singularities may be avoided if matter sources possess a de Sitter core at the center of spacetime. Building on this idea, Bardeen proposed the first static, spherically symmetric, regular black hole solution \cite{jm68}, now called the Bardeen. Nearly three decades after Bardeen’s proposal, Ayón-Beato and García \cite{ea98, ea99, ea99a, ea00, ea05} offered the first field-theoretic interpretation of the Bardeen black hole. They proposed that the solution could arise from Einstein’s field equations with a nonlinear electrodynamics source, specifically a magnetic monopole. Indeed, the definition of a regular black hole can be understood from two perspectives. The first is a coordinate-independent criterion, which requires the finiteness of curvature invariants \cite{id92, ea98, ka01}. In this view, a regular black hole is described as a black hole spacetime in which all curvature invariants remain finite, a notion closely connected to Markov’s limited curvature conjecture \cite{ma82}. The second is a coordinate-dependent criterion, based on geodesic completeness \cite{rc20a, rc22}, wherein a regular black hole is defined as a spacetime admitting complete null and timelike geodesics. These two definitions are not generally equivalent, implying that certain models may fulfill one condition while violating the other. \cite{rp68, gj15}. Recently, numerous researchers introduced several regular black hole solutions in the literature \cite{ab94, cb96, ab00, ac99, sa06, cb13, sg15, bt14, ma14, id15}. 

The black holes indicate a dark shadow, the photon ring, and the relativistic images produced from the gravitational lensing effect on the event horizon when numerous photons pass near a black hole. Studies revealed that the black hole shadow resulting from the effect of gravitational lensing \cite{H. Falcke, C. Bambi, K. Hioki, L. Amarilla, A. Abdujabbarov, P.V.P, R. Ghosh}. The important effect of gravitational lensing may be discussed under the strong field limit. Darwin \cite{Darwin} has investigated the strong field limit. subsequently, in another study, the authors obtained the gravitational lens equation in the strong field limit, and the gravitational lens of a Schwarzschild black hole was investigated \cite{K. S.}. Latterly, the same authors investigate the naked singularity lens and relativistic images of Schwarzschild black hole lensing, respectively \cite{70,71,72}. In another study, Bozza et al. \cite{Bozza} provide the analytical method to calculate the strong deflection angle of the spherically symmetric black hole, and investigated the observable of a spherically symmetric black hole based on the lens equation. 

In the current study, our main objective is to investigate weak gravitational lensing using the Gauss-Bonnet theorem (GBT), which is also known as the Gibbons-Werner method (GWM). Black hole holes may not be observed owing to it's strong gravitational force, there is another way to estimate their existence in our universe through the study of geodesic equations of light rays in a curved spacetime geometry. The curved spacetime is due to the presence of black holes, hence, by this method, one may get the information about their nature and valuable information from the black holes. To study the nature of black holes, gravitational lensing is an interesting scientific technique; hence, for most cases, the strong lensing regime is required. To detect the ultra compact objects such as boson stars or other exotic objects, the strong lensing gives more information by using the experimental method \cite{6}. Recently, scientists detected the black hole horizon through the Event Horizon Telescope \cite{7,8}. This topic has drawn the great interest of researchers after the detection of gravitational waves and the horizon of black holes. 

Gibbons and Werner \cite{gibbons1} introduced the unique technique to calculate the deflection angle, such as they have used the Gauss Bonnet Theorem in optical geometry. 
Physical significance depends on the fact that one may observe the bending of a light ray as a global effect. Weak gravitational lensing uses the fine property of the differential deflection shown by the bending of light in order to investigate the structures of the cosmos. For achieving these results, the deflection angle is investigated by considering the optical geometry in the context of GBT given by \cite{gibbons1}\\
\begin{equation}\label{GBT}
\iint\limits_{D}\mathcal{G}\,\mathrm{d}\mathcal{A}+\oint\limits_{\partial D}\kappa \,\mathrm{d}t+\sum_{i}\theta _{i}=2\pi \chi (D),
\end{equation}
where $\mathcal{G}$ is the Gaussian curvature, $\kappa$ is the geodesic curvature, $\theta_{i}$ is the exterior angle at the $i^{th}$ vertex, $\mathrm{d}\mathcal{A}$ is the surface element, and $\chi$ is the Euler characteristic element. Here, $\chi (D) = 1$, as $D$ is a non-singular region. Various studies have been conducted by using the GBT method on black holes, wormholes and by considering different spacetimes \cite{Werner, Övgün1, Övgün2, Javed, Takizawa, Ishihara, Pantig, Pantig2, Xie, Kumaran}. 

The shadow of the black hole is described as a critical curve interior that divides the capture orbits, which spiral into the black hole, from those veering away that is indicating that photons enter the black hole or escape. Although the shadow size of black hole is basically depend over the intrinsic parameters of black hole, and it's contour is calculated by the orbital instability of light rays from the photon, it looks a dark, 2-dimensional disc for the distant observer \cite{Pantig2, Övgün3,Okyay,Allahyari,Roy,Vagnozzi,Khodadi,Khodadi2,Kumar,Rahaman,Belhaj,Sun}. In this article, we also investigate the shadow cast by the present black hole with respect to the impact $q$ and mass $m$, which ensures that the shadow region shrinks for increasing values of $q$ and expands for increasing values of $m$. The pioneering work on the shadow of a black hole was conducted by Bardeen \cite{J.M}. After that, numerous studies have been conducted in order to investigate the shadow of a black hole by considering different techniques, for example, some consider electric charge on black hole, whereas some assume magnetic charge and accretion flow, etc \cite{Kumaran, P.V.P, H.C.D.L., Y.Meng, R.Ling, M.Wang, Z.Hu, W.D, Jusufi, Gao}. From the literature, it is predicted that the shadow of spherical symmetric black holes are circular \cite{J.L,J.P}, while spinning black holes deformed the shadow \cite{S.,H.}. Bardeen also calculated the angular radius for the Schwarzschild black holes \cite{J.M}, latterly, by introducing two new observable parameters were given in Ref. \cite{Hioki}. The above-mentioned investigation sparked the extensive studies into the shadow of black holes in different considerations, namely, Kerr–Newman, Reissner–Nordstrom, Kerr–Sen, Kerr–Taub-NUT, braneworld, non-Kerr,
higher dimensional and regular black holes \cite{A.F,A.F2,A.de,R.Takahashi,K.Hioki,A.Abd2,A.Gre,F.Atam,L.Ama,M.Amir,U.Pap}. Some authors have investigated the shadow and photon ring of astrophysical black hole by considering the luminous accretion flow around the black hole. In one of the pioneering studies, the shadow and photon ring observation characteristics are investigated, depending on the position and nature of accretion flow \cite{J.P.Lum}. In this study, the author considered the spherical symmetric black hole surrounded by a thin accretion disc. In another study, Bambi investigated the difference between black holes and wormholes \cite{C.Bambi1}. Gralla et al. \cite{S.E} considered the Schwarzschild black hole to investigate the simple case of emission from geometrically and optically thin disc and predicted the bright ring close to the Schwarzschild black hole shadow comprises the lensed ring, the direct emission, and the photon ring, that are calculated from the times of the intersects between the thin accretion disc and a light ray. In another study, Naryan et al. investigated the spherical symmetric model of optically thin accretion disc around Schwarzschild black hole and demonstrated that the accretion flow hardly influenced the shadow of the black hole \cite{R.Nara}. 

The current article is structured as follows: We have formulated the regular black hole solution coupled with nonlinear electrodynamics and explored its thermodynamic features in Sec.- \ref{secII}. The Gauss-Bonnet method is described in Sec.- \ref{secIII}. We have estimated the deflection angle in vacuum medium in See.- \ref{secIV} and analyzed the obtained deflection angle with respect to the impact parameter $b$, charge parameter $q$, and black hole mass $m$ in See.- \ref{secV}. In See.- \ref{secVI}, we have calculated the deflection angle in the plasma medium and analyzed it graphically in See.- \ref{secVII}.  The Sec.-\ref{secVIII} addresses the time delay in the spacetime of the present black hole. We have analyzed the shadow cast and greybody bound of the black hole in  Sec.-\ref{secIX} and Sec.-\ref{secX}, respectively. Finally, the results and conclusion have been derived in See.-\ref{secXI}.

%%%%%%%%%%%%%%%%%%%%%%%%%%%%%%%%%%%%%%%%%%%%%%%%%%%%%%%%%%%%%%%%%%
\section{Regular Black Hole Coupled with Nonlinear Electrodynamics }\label{secII}
%%%%%%%%%%%%%%%%%%%%%%%%%%%%%%%%%%%%%%%%%%%%%%%%%%%%%%%%%%%%%%%%%%
We commence with a brief discussion on the regular black hole solution of the Einstein field equations in the framework of standard general relativity, following the seminal work of Eloy Ayón–Beato et al. \cite{ea98}.  The action formulated within the framework of Einstein–dual nonlinear electrodynamics can be expressed as \cite{hs87} 
\begin{eqnarray}
    \mathcal{S} = \int dv\left(\frac{1}{16\pi}R-\frac{1}{4\pi}\mathcal{L}(F)\right),\label{S}
\end{eqnarray}
where $R$ denotes the Ricci scalar, and $\mathcal{L}$ is a funstion of $F = \frac{1}{4}F_{\mu\nu}F^{\mu\nu}$. The system under consideration can also be reformulated in terms of another function derived through a Legendre transformation \cite{hs87}
\begin{eqnarray}
    \mathcal{H}\equiv 2F \mathcal{L}_F-\mathcal{L}.
\end{eqnarray}
Defining $P_{\mu\nu} \equiv \mathcal{L}_FF_{\mu\nu}$, one finds that $\mathcal{H}$ is a function $P \equiv \frac{1}{4}P_{\mu\nu}P^{\mu\nu}=(\mathcal{L}_F)^2F$ such that $d\mathcal{H} = (\mathcal{L}_F)^{-1}d((\mathcal{L}_F)^2F) = \mathcal{H}_P dP$. In terms of $\mathcal{H}$, the nonlinear electromagnetic Lagrangian appearing in the action (\ref{S}) can be expressed as $\mathcal{L} = 2P\mathcal{H}_P - \mathcal{H}$, which depends explicitly on the antisymmetric tensor $P_{\mu\nu}$.  The particular choice of the function $\mathcal{H}$, specifying the nonlinear electrodynamic source employed, is given by
\begin{eqnarray}
    \mathcal{H}(P) = P\frac{1-3\sqrt{-2q^2P}}{\left(1+\sqrt{-2q^2P}\right)^3}-\frac{3}{2q^2 s}\left(\frac{\sqrt{-2q^2P}}{1+\sqrt{-2q^2P}}\right)^{5/2},\label{H}
\end{eqnarray}
where $s=|q|/2m$  and the invariant $P$ is a negative quantity. The corresponding Lagrangian is then obtained as
\begin{eqnarray}
    \mathcal{L} = P\frac{\left(1-8\sqrt{-2q^2P}-6q^2P\right)}{\left(1+\sqrt{-2q^2P}\right)^4}-\frac{3}{4q^2 s}\frac{\left(-2q^2P\right)^{5/4}\left(3-2\sqrt{-2q^2P}\right)}{\left(1+\sqrt{-2q^2P}\right)^{7/2}}.
\end{eqnarray}
The function (\ref{H}) satisfies the essential requirements for a viable nonlinear electromagnetic model: (i) correspondence with Maxwell's theory, i.e., $\mathcal{H} \approx P$ in the weak-field limit ($P \ll 1$), and (ii) compliance with the weak energy condition, which demands $\mathcal{H} < 0$ and $\mathcal{H}P > 0$. It is worth emphasizing that the solution, besides being regular and satisfying the weak energy condition, possesses an additional noteworthy property: it does not admit a Cauchy surface. Consequently, it does not conflict with the Penrose singularity theorem, whose assumptions include the validity of the null energy condition, the existence of a noncompact Cauchy surface, and the presence of a closed trapped surface, leading to the conclusion of null geodesic incompleteness of spacetime.

Now, the Einstein and nonlinear electrodynamic field equations for the action (\ref{S}) can be expressed as
\begin{eqnarray}
    &&G_\mu^{~\nu} = 2\left(\mathcal{H}_PP_{\mu\lambda}P^{\nu\lambda}-\delta_\mu^{~\nu}(2P\mathcal{H}_P-\mathcal{H})\right),\label{est}
    \\
    &&\nabla_\mu P^{\alpha\mu} = 0.\label{nabla}
\end{eqnarray}

To obtain a singularity-free, static, and spherically symmetric black hole solution coupled to nonlinear electrodynamics and satisfying the weak energy condition, we adopt the static and spherically symmetric ansatz
\begin{eqnarray}
    ds^2 = -\left(1-\frac{2m}{r}+\frac{Q(r)}{r^2}\right)dt^2+\left(1-\frac{2m}{r}+\frac{Q(r)}{r^2}\right)^{-1}dr^2+r^2(d\theta^2+\sin^2\theta d\phi^2)\label{ds}
\end{eqnarray}
together with the following ansatz for the antisymmetric field $P_{\mu\nu} = 2\delta^t_{[\mu}\delta^r_{\nu]}D(r)$. Here, $m$ is the mass of the black hole. With these choices, the equation (\ref{nabla}) yields
\begin{eqnarray}
    P_{\mu\nu} = 2\delta^t_{[\mu}\delta^r_{\nu]}\frac{q}{r^2} \rightarrow P = -\frac{D^2}{2}=-\frac{q^2}{2r^4},
\end{eqnarray}
where the integration constant has been chosen as $q$, it effectively represents the electric charge of the black hole solution.

The $\,^{~t}_{t}$ component of the Einstein equations (\ref{est}) yields the fundamental relation
\begin{eqnarray}
    \frac{rQ'-Q}{r^4} = 2\mathcal{H}(P)\label{nn}.
\end{eqnarray}

On using $\mathcal{H}$ from Eq. (\ref{H}), with $P = -\tfrac{q^2}{2r^4}$, the integral of the above result (\ref{nn}) can be expressed as
\begin{eqnarray}
    Q = q^2r\int_r^\infty dy \left(\frac{6my^2}{\left(y^2+q^2\right)^{5/2}}+\frac{y^2(y^2-3q^2)}{\left(y^2+q^2\right)^3}\right), 
\end{eqnarray}
 Thus, one finally arrives at the following expression
 \begin{eqnarray}
     Q = 2mr - \frac{2mr^4}{\left(r^2+q^2\right)^{3/2}}+\frac{q^2r^4}{\left(r^2+q^2\right)^{2}}.
 \end{eqnarray}
By substituting $Q(r)$ into Eq.~(\ref{ds}), we obtain the metric ansatz for a singularity-free, static, and spherically symmetric black hole solution, coupled to nonlinear electrodynamics and satisfying the weak energy condition, within the framework of general relativity as
\begin{equation}
    ds^2 = -\left(1-\frac{2 m r^2}{\left(q^2+r^2\right)^{3/2}}+\frac{q^2 r^2}{\left(q^2+r^2\right)^2}\right)dt^2+\left(1-\frac{2 m r^2}{\left(q^2+r^2\right)^{3/2}}+\frac{q^2 r^2}{\left(q^2+r^2\right)^2}\right)^{-1}dr^2+r^2(d\theta^2+\sin^2\theta d\phi^2).\label{bh}
\end{equation}

 It is noted that the above black hole solution reduces to the Schwarzschild black hole whenever $q = 0$.  Furthermore, the evaluation of the electric field, defined as $E = F_{tr} = \mathcal{H}_P D$, by employing expression (\ref{H}) for $\mathcal{H}$, yields the electric field associated with the black hole solution (\ref{bh}) as
\begin{equation}
    E = qr^4\left(\frac{r^2-5q^2}{\left(r^2+q^2\right)^2}+\frac{15m}{2\left(r^2+q^2\right)^{7/2}}\right).
\end{equation}

Now, the metric coefficient $-g_{tt}$ of the above black hole solution (\ref{bh}) can be expressed as
\begin{eqnarray}
    -g_{tt} = 1-\frac{2 m r^2}{\left(q^2+r^2\right)^{3/2}}+\frac{q^2 r^2}{\left(q^2+r^2\right)^2} = 1-\frac{2m}{q}\frac{ (r/q)^2}{\left(1+(r/q)^2\right)^{3/2}}+\frac{(r/q)^2}{\left(1+(r/q)^2\right)^2}
\end{eqnarray}

\begin{figure}[!htbp]
\begin{center}
\begin{tabular}{rl}
\includegraphics[width=8cm]{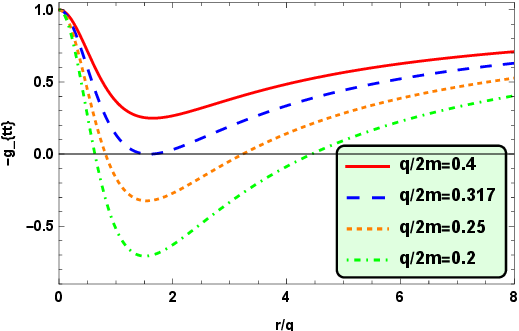}
\end{tabular}
\end{center}
\caption{Behavior of -$g_{tt}$ against $r/q$ for different values of $q/2m$.}\label{fig1a}
\end{figure}

The profile of $-g_{tt}$ is depicted in Fig. \ref{fig1a} that ensures that the metric (\ref{bh}) is a black hole for a certain range of the mass and charge as $q/2m \leq 0.317$ or $q \leq 0.635m \approx 0.6m$ (See details in Ref. \cite{ea98}). Indeed, the inner event horizon $r_-$ and outer event horizon $r_+$ exist whenever $q < 0.6m$. Following it, we have provided the values of inner and outer event horizons of the present black hole in Table- \ref{tab1} for different values of $q$ and $m$.  Moreover,  the Kretschmann scalar of the  black hole solution (\ref{bh}) is obtained as 
\begin{eqnarray}
    \mathcal{K}_s &=& \frac{4}{\left(q^2+r^2\right)^8} \bigg[3 m^2 \left(q^2+r^2\right) \left(8 q^8-4 q^6 r^2+47 q^4 r^4-12 q^2 r^6+4 r^8\right)-6 m q^2 \big(4 q^8-5 q^6 r^2+32 q^4 r^4-15 q^2 r^6\nonumber
    \\
    &&+4 r^8\big) \sqrt{q^2+r^2}+2 q^4 \left(3 q^8-6 q^6 r^2+34 q^4 r^4-22 q^2 r^6+7 r^8\right)\bigg].\label{K}
\end{eqnarray}

%%%%%%%%%%%%%%%%%%%%%%%%%%%%%%%%%%%%%%%%%%%%%%%%%%%%%%%%%%%%%%%%%%%%%
%%%%%%%%%%%%%%%%%%%%%%%%%%%%%%%%%%%%%%%%%%%%%%%%%%%%%%%%%%

The above result of the Kretschmann scalar confirms that the black hole is regular everywhere. It is worth noting that the Kretschmann scalar (\ref{K}) reduces to $48m^{2}/r^{6}$ for $q = 0$, which coincides with the Kretschmann scalar of the Schwarzschild black hole. In this context, we also determine the non-vanishing Ricci Tensor for the black hole (\ref{bh}), obtained as 
\begin{eqnarray}
    R_{tt} &=& \frac{q^2 \left[\sqrt{r^2+q^2} \left(r^4+q^4+3q^2r^2\right)-2mr^2\left(r^2+q^2\right)\right] \left[m \left(9 r^4+3 q^2 r^2-6 q^4\right)+\left(r^4+3 q^4-8 q^2 r^2\right) \sqrt{r^2+q^2}\right]}{\left(r^2+q^2\right)^7},\nonumber
    \\
    \\
    R_{rr} &=& \frac{q^2 \left[m \left(6 q^4-3 q^2 r^2-9 r^4\right)-\left(3 q^4-8 q^2 r^2+r^4\right) \sqrt{r^2+q^2}\right]}{\left(r^2+q^2\right)^{5/2} \left(r^4+3 q^2 r^2+q^4\right)-2mr^2 \left(r^2+q^2\right)^3},
    \\
    R_{\theta\theta} &=& \frac{q^2 r^2 \left[3q^2-6 m \sqrt{r^2+q^2}-r^2\right]}{\left(r^2+q^2\right)^3},
    \\
    R_{\phi\phi} &=& \frac{q^2 r^2 \left[3q^2-6 m \sqrt{r^2+q^2}-r^2\right]\sin^2\theta}{\left(r^2+q^2\right)^3}.
\end{eqnarray}

and the Ricci scalar is obtained as
\begin{equation}
R = \frac{2 q^2 r^2 \left[10q^2-15 m \sqrt{r^2+q^2}-2 r^2\right]}{\left(r^2+q^2\right)^4}.
\end{equation}

 All the findings support the regular behavior of the considered black hole (\ref{bh}) throughout the spacetime. In the absence of electric charge ($q = 0$), we recover $R_{tt} = R_{rr} = R_{\theta\theta} = R_{\phi\phi} = R = 0$, as expected for the Schwarzschild black hole.

We now study some thermodynamic features of the present black hole solution. The condition $g_{tt}(r_\pm) = 0$ at the outer horizon yields the mass of the black hole as
\begin{eqnarray}
    m(r_\pm) = \frac{q^4+3 q^2 r_\pm^2+r_\pm^4}{2 r_\pm^2 \sqrt{q^2+r_\pm^2}}.
\end{eqnarray}

 The surface gravity of the  black hole is obtained as
 \begin{eqnarray}
    \kappa(r_+) = \frac{1}{2}\left[\frac{d g_{tt}}{dr}\right]_{r_+} = \frac{r_+}{\left(q^2+r_+^2\right)^{7/2}} \left[m \left(r_+^4-q^2 r_+^2-2 q^4\right)+q^2 \left(q^2-r_+^2\right) \sqrt{q^2+r_+^2}\right].
\end{eqnarray}
 
The Hawking temperature of the black hole is obtained as
\begin{eqnarray}
    T(r_+) = \frac{1}{4\pi}\left[\frac{d g_{tt}}{dr}\frac{1}{\sqrt{-g_{tt}g_{rr}}}\right]_{r_+} = \frac{r_+}{2\pi\left(q^2+r_+^2\right)^{7/2}} \left[m \left(r_+^4-q^2 r_+^2-2 q^4\right)+q^2 \left(q^2-r_+^2\right) \sqrt{q^2+r_+^2}\right].
\end{eqnarray}

Also, the heat capacity $C$, entropy $S$, and Gibbs free energy $F$ are obtained as
\begin{eqnarray}
    C &=& \frac{dm(r_+)}{dT(r_+)} = \frac{2 \pi  \left(q^2+r_+^2\right)^{5/2} \left(2 q^6+3 q^4 r_+^2+q^2 r_+^4-r_+^6\right)}{r_+^9-8 q^2 r_+^7-12 q^4 r_+^5-11 q^6 r_+^3-2 q^8 r_+},
\\
    S &=& \int \frac{1}{T(r_+)}\frac{dm(r_+)}{dr_+}dr_+ = \frac{\pi}{r_+}  \left(r_+^2-2 q^2\right) \sqrt{q^2+r_+^2}-3 \pi  q^2 \log \left(\sqrt{q^2+r_+^2}-r_+\right),
\\
    F &=& m(r_+) -T(r_+)S(r_+)\nonumber
    \\
    &=& \frac{\sqrt{q^2+r_+^2} \left(r_+^8+13 q^2 r_+^6+17 q^4 r_+^4+6 q^6 r_+^2-2 q^8\right)-3 \left(2 q^8 r_++3 q^6 r_+^3+q^4 r_+^5-q^2 r_+^7\right) \log \left(\sqrt{q^2+r_+^2}-r_+\right)}{4 r^2 \left(q^2+r_+^2\right)^3}.\nonumber
    \\
\end{eqnarray}

It is noted that all the present thermodynamic features become of the Schwarzchild black hole whenever the charge parameter $q = 0$. Now, we are going to estimate the deflection angle of light in the weak field approximation of the present black hole (\ref{bh}) using the Gauss-Bonnet method within the non-plasma and plasma medium.

%%%%%%%%%%%%%%%%%%%%%%%%%%%%%%%%%%%%%%%%%%%%%%%%%%%%%%%%%%%%%%%%%%%%%
\begin{figure}[!htbp]
\begin{center}
\begin{tabular}{rl}
\includegraphics[width=8cm]{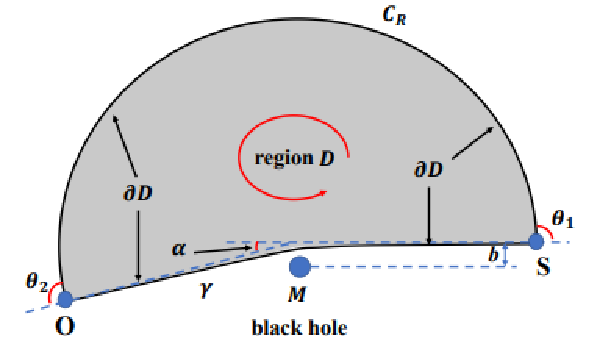}
\end{tabular}
\end{center}
\caption{This figure illustrates the chosen region $D$ on the equatorial plane of the optical manifold in an asymptotically flat spacetime, which facilitates the application of the Gauss-Bonnet method as expressed in Eq.~(\ref{GBT}) for calculating the gravitational deflection angle.}\label{fig1b}
\end{figure}
%%%%%%%%%%%%%%%%%%%%%%%%%%%%%%%%%%%%%%%%%%%%%%%%%%%%%%%%%%

%%%%%%%%%%%%%%%%%%%%%%%%%%%%%%%%%%%%%%%%%%%%%%%%%%%%%%%%%%%%%%%%%%
\section{Gauss-Bonnet method}\label{secIII}
%%%%%%%%%%%%%%%%%%%%%%%%%%%%%%%%%%%%%%%%%%%%%%%%%%%%%%%%%%%%%%%%%%
Recently, Gibbons and Werner \cite{gibbons1} introduced a very useful method for estimating the weak deflection angle of light in the gravitational field of a massive astrophysical object that has asymptotic behaviour. Indeed, Gibbons and Werner's theorem described the  connection between the intrinsic geometry of the spacetime and its topology of the non-singular region $D$  surrounded by the beam of light with boundary $\partial D$, stated as \cite{gibbons1}
\begin{equation}\label{GBT}
\iint\limits_{D}\mathcal{G}\,\mathrm{d}\mathcal{A}+\oint\limits_{\partial D}\kappa \,\mathrm{d}t+\sum_{i}\theta _{i}=2\pi \chi (D),
\end{equation}
where $\mathcal{G}$ is the Gaussian curvature, $\kappa$ is the geodesic curvature, $\theta_{i}$ is the exterior angle at the $i^{th}$ vertex, $\mathrm{d}\mathcal{A}$ is the surface element, and $\chi$ is the Euler characteristic element. Here, $\chi (D) = 1$, as $D$ is a non-singular region.

In evaluating gravitational deflection angle, the region $D$ in the Gauss-Bonnet theorem is defined as follows: For an asymptotically flat spacetime, when both the source $S$ and the observer $O$ are located far from the central black hole, $D$ is chosen as a simply connected region in the equatorial plane, bounded by $\partial D$. The boundary $\partial D$ consists of two parts: the particle trajectory $\gamma$ from $S$ to $O$, and a circular arc $C_R$ connecting $O$ and $S$. The central black hole lies outside the region $D$, thereby excluding any spacetime singularities from $D$. This construction is schematically depicted in Fig. \ref{fig1b}.

The mathematical formula for computing the geodesic curvature is given as
\begin{equation}
\kappa =g^{\text{op}}\,\left(\nabla _{\dot{%
\gamma}}\dot{\gamma},\ddot{\gamma}\right)~~\text{satisfying} ~~ g^{op}(\dot{\gamma},\dot{%
\gamma}) = 1,
\end{equation}
where $\ddot{\gamma}$ denotes the unit acceleration vector. For $R\rightarrow \infty$, the jump angles $\theta_1$ and $\theta_2$ to the source and viewer, respectively become $\pi/2$ i.e. $\theta _1$ + $ \theta _2\rightarrow \pi$  \cite{gibbons1}. Moreover, $\kappa(\gamma _{g^{op}})=0$ for a geodesic $\gamma _{g^{op}}$. Consequently, the geodesic curvature  reads as 
\begin{equation}
\kappa (\gamma_{R})=|\nabla _{\dot{\gamma}_{R}}\dot{\gamma}_{R}|,
\end{equation}

The geodesic curvature's radial part can be stated as
\begin{equation}
\left( \nabla _{\dot{\gamma}_{R}}\dot{\gamma}_{R}\right) ^{r}=\dot{\gamma}_{R}^{\varphi
}\,\left( \partial _{\varphi }\dot{\gamma}_{R}^{r}\right) +\tilde{\Gamma} _{\varphi
\varphi }^{r}\left( \dot{\gamma}_{R}^{\varphi }\right) ^{2}, \label{12}
\end{equation}
with the condition $\gamma_{R}:=r(\varphi)=R=\text{constant}$ for large $R$. Here,  $\tilde{\Gamma}_{\varphi\varphi }^{r}$ represents the Christoffel symbol of the optical metric. It is evident from the preceding equation that the first term will vanish as the topological effect is not involved and the second term can be derived using $\tilde{g}_{\varphi \varphi}\dot{\gamma}_{R}^{\varphi } \dot{\gamma}_{R}^{\varphi }=1 $, known as the unit speed condition. Therefore, the geodesic curvature becomes as
\begin{eqnarray}\notag\label{gcurvature}
\lim_{R\rightarrow \infty }\kappa (\gamma_{R}) &=&\lim_{R\rightarrow \infty
}\left\vert \nabla _{\dot{\gamma}_{R}}\dot{\gamma}_{R}\right\vert  \notag 
\rightarrow \frac{1}{R}.
\end{eqnarray}%

%%%%%%%%%%%%%%%%%%%%%%%%%%%%%%%%%%%%%%%%%%%%%%%%%%%%%%%%%%%%%%%%%%%%%
\begin{figure}[!htbp]
\begin{center}
\begin{tabular}{rl}
\includegraphics[width=8cm]{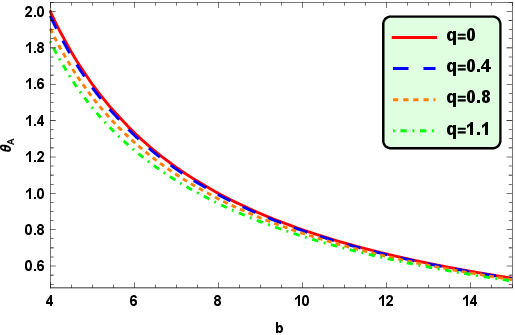}
\end{tabular}
\begin{tabular}{rl}
\includegraphics[width=8cm]{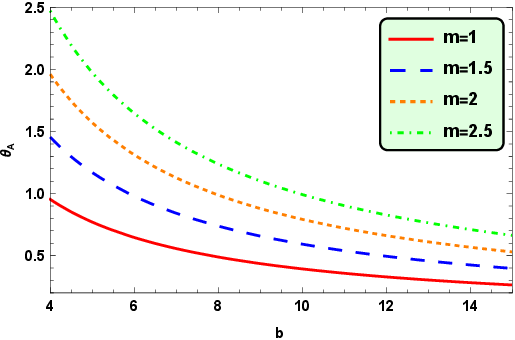}
\end{tabular}
\end{center}
\caption{Deflection angle $\theta_A$ against the impact parameter $b$ corresponding to $m = 2$ (Left) and $q = 0.5$ (Right).}\label{fig1}
\end{figure}
%%%%%%%%%%%%%%%%%%%%%%%%%%%%%%%%%%%%%%%%%%%%%%%%%%%%%%%%%%

%%%%%%%%%%%%%%%%%%%%%%%%%%%%%%%%%%%%%%%%%%%%%%%%%%%%%%%%%%%%%%%%%%%%%
\begin{figure}[!htbp]
\begin{center}
\begin{tabular}{rl}
\includegraphics[width=8cm]{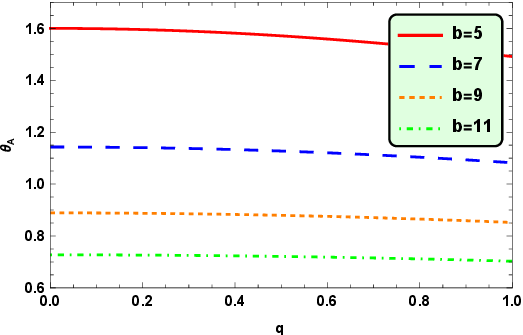}
\end{tabular}
\begin{tabular}{rl}
\includegraphics[width=8cm]{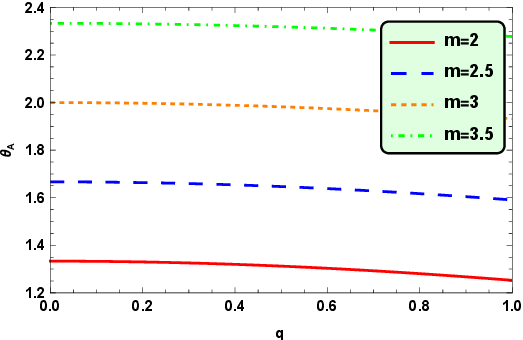}
\end{tabular}
\end{center}
\caption{Deflection angle $\theta_A$ against the charge paramete $q$ corresponding to $m = 2$ (Left) and $b = 6$ (Right).}\label{fig2}
\end{figure}
%%%%%%%%%%%%%%%%%%%%%%%%%%%%%%%%%%%%%%%%%%%%%%%%%%%%%%%%%%

Subsequently, we arrived at  $\kappa (\gamma_{R}) dt$ = $d\phi$ for the significantly large $R$. Therefore, the Gauss-Bonnet theorem (\ref{GBT}) reads as
\begin{equation}
\iint\limits_{D}\mathcal{G}\,\mathrm{d}\mathcal{A}+\oint\limits_{\gamma_{R}}\kappa \,%
\mathrm{d}t\overset{{R\rightarrow \infty }}{=}\iint\limits_{D%
_{\infty }}\mathcal{G}\,\mathrm{d}\mathcal{A}+\int\limits_{0}^{\pi + \theta_A}\mathrm{d}\varphi
=\pi,
\end{equation}

Now, we consider the light beam which follows a straight line approximation  $r_\gamma = b/ \sin\phi$, where $b$ is the impact parameter to estimate the weak deflection angle. Finally, the deflection angle is obtained in the following form
\begin{eqnarray}\label{GBT2}
\theta_A &=&-\int\limits_{0}^{\pi}\int\limits_{b/ \sin\phi}^{\infty} \mathcal{G} \mathrm{d}\mathcal{A}.\label{da}
\end{eqnarray}
It is noted that the impact parameter $b$ can be approximated by the closest distance of approach to the black hole in a first-order approximation.

%%%%%%%%%%%%%%%%%%%%%%%%%%%%%%%%%%%%%%%%%%%%%%%%%%%%%%%%%%%%%%%%%%
\section{Deflection angle in vacuum medium}\label{secIV}
%%%%%%%%%%%%%%%%%%%%%%%%%%%%%%%%%%%%%%%%%%%%%%%%%%%%%%%%%%%%%%%%%%
 We consider the null geodesics in the gravitational field of the present black hole (\ref{bh}) to calculate the Gaussian optical curvature that can be expressed from the null geodesic equations $ds^2$ = 0 at the equatorial plane $\theta = \pi/2$ as 
\begin{equation}
dt^2=\left[1-\frac{2 m r^2}{\left(q^2+r^2\right)^{3/2}}+\frac{q^2 r^2}{\left(q^2+r^2\right)^2}\right]^{-2}dr^2+r^2\left[1-\frac{2 m r^2}{\left(q^2+r^2\right)^{3/2}}+\frac{q^2 r^2}{\left(q^2+r^2\right)^2}\right]^{-1} d\phi^2.\label{bh1}
\end{equation}

Also, the optical metric can be expressed with a new function  $f(r^\star)$, where $r^\star$ is the  Regge-Wheeler tortoise static radial coordinate,  in the following form
\begin{equation}
\mathrm{d}t^2 \equiv g_{ab}^{op} \mathrm{d}x^a \mathrm{d}x^b={\mathrm{d}r^{\star}}^{2}+{f(r^{\star})}^2 \mathrm{d}\varphi^2.\label{rw}
\end{equation}

%%%%%%%%%%%%%%%%%%%%%%%%%%%%%%%%%%%%%%%%%%%%%%%%%%%%%%%%%%%%%%%%%%%%%
\begin{figure}[!htbp]
\begin{center}
\begin{tabular}{rl}
\includegraphics[width=8cm]{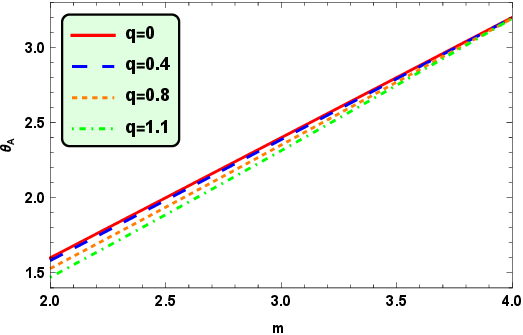}
\end{tabular}
\begin{tabular}{rl}
\includegraphics[width=8cm]{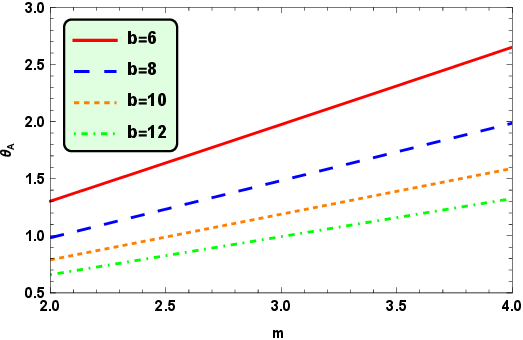}
\end{tabular}
\end{center}
\caption{Deflection angle $\theta_A$ against the mass $m$ corresponding to $b = 5$ (Left) and $q = 0.6$ (Right).}\label{fig3}
\end{figure}
%%%%%%%%%%%%%%%%%%%%%%%%%%%%%%%%%%%%%%%%%%%%%%%%%%%%%%%%%%

 Now, Eqs. (\ref{bh1}) and (\ref{rw}) simultaneously yield the following results
\begin{eqnarray}
\mathrm{d}r^\star &=&\left[1-\frac{2 m r^2}{\left(q^2+r^2\right)^{3/2}}+\frac{q^2 r^2}{\left(q^2+r^2\right)^2}\right]^{-1}dr,
\end{eqnarray}
\begin{eqnarray}
f(r^\star)&=& r \left[1-\frac{2 m r^2}{\left(q^2+r^2\right)^{3/2}}+\frac{q^2 r^2}{\left(q^2+r^2\right)^2}\right]^{-\frac{1}{2}}.
\end{eqnarray}

The mathematical formula for Gaussian curvature $\mathcal{G}$ of the optical surface is given as \cite{gibbons1}
\begin{eqnarray}
\mathcal{G}&=&-\frac{1}{f(r^{\star })}\frac{\mathrm{d}^{2}f(r^{\star })}{\mathrm{d}{%
r^{\star }}^{2}}= -\frac{1}{f(r^{\star })}\left[ \frac{\mathrm{d}r}{\mathrm{d}r^{\star }}%
\frac{\mathrm{d}}{\mathrm{d}r}\left( \frac{\mathrm{d}r}{\mathrm{d}r^{\star }}%
\right) \frac{\mathrm{d}f}{\mathrm{d}r}+\left( \frac{\mathrm{d}r}{\mathrm{d}%
r^{\star }}\right) ^{2}\frac{\mathrm{d}^{2}f}{\mathrm{d}r^{2}}\right].\label{g}
\end{eqnarray}

On using the above formula (\ref{g}),  we get the optical Gaussian curvature for the optical black hole metric (\ref{bh1}) in the following form
\begin{eqnarray}
\mathcal{G} &=& -\frac{2 m}{r^3}+\frac{3 m^2+3 q^2}{r^4}+\frac{12 m q^2}{r^5}-\frac{33 m^2 q^2+18 q^4}{r^6}-\frac{9 m q^4}{4 r^7}+O\left(\frac{1}{r^8}\right).\label{cur}
\end{eqnarray}

Now, the surface element $d\mathcal{A}$ can be approximated as
\begin{eqnarray}
    d\mathcal{A} = \sqrt{g^{opt}}dr d\phi \simeq r dr d\phi.
\end{eqnarray}

Therefore, from Eq. (\ref{da}), we obtain the weak deflection angle of light in the non-plasma medium as
\begin{eqnarray}
\theta_A \simeq \frac{4 m}{b} -\frac{3 \pi  q^2}{4 b^2}-\frac{16 m q^2}{3 b^3}+\frac{9 \pi  q^2 \left(11 m^2+6 q^2\right)}{32 b^4}+\frac{12 m q^4}{25 b^5}.\label{DA}
\end{eqnarray}

Thus, it is immediately apparent that the charge parameter $q$ influences the weak deflection angle of the present black hole. For $q = 0$, one can see that the deflection angle becomes $\theta_A = 4 m/b$, which is the deflection angle of light in the weak gravitational field of the Schwarzschild black hole \cite{jb03}. Now, we are willing to examine the behaviours of the deflection angle graphically with respect to the impact parameter, charge parameter and black hole mass.

%%%%%%%%%%%%%%%%%%%%%%%%%%%%%%%%%%%%%%%%%%%%%%%%%%%%%%%%%%%%%%%%%%%%%
\begin{figure}[!htbp]
\begin{center}
\begin{tabular}{rl}
\includegraphics[width=8cm]{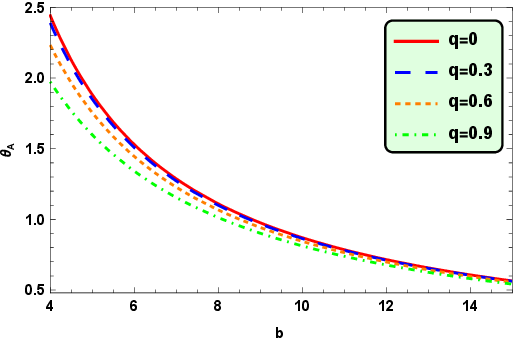}
\end{tabular}
\begin{tabular}{rl}
\includegraphics[width=8cm]{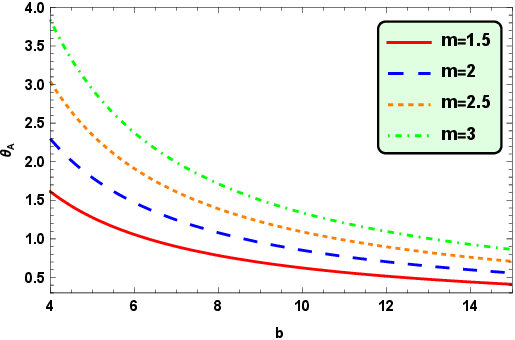}
\end{tabular}
\end{center}
\caption{Deflection angle $\theta_A$ against the impact parameter $b$ corresponding to $m = 2$ (Left) and $q = 0.5$ (Right).}\label{fig4}
\end{figure}
%%%%%%%%%%%%%%%%%%%%%%%%%%%%%%%%%%%%%%%%%%%%%%%%%%%%%%%%%%

%%%%%%%%%%%%%%%%%%%%%%%%%%%%%%%%%%%%%%%%%%%%%%%%%%%%%%%%%%%%%%%%%%
\section{GRAPHICAL ANALYSIS of Deflection angle FOR vacuum  MEDIUM}\label{secV}
%%%%%%%%%%%%%%%%%%%%%%%%%%%%%%%%%%%%%%%%%%%%%%%%%%%%%%%%%%%%%%%%%%

 To examine the effect of the impact parameter $b$, charge parameter $q$,  and the black hole mass $m$ on the deflection angle we demonstrate the deflection angle graphically. At first, we plot the deflection angle $\theta_A$ against the impact parameter $b$ by varying the charge parameter $q$,  and the black hole mass $m$. Second, we plot the deflection angle $\theta_A$ against the charge parameter $q$ by varying the impact parameter $b$,  and the black hole mass $m$. At the end,  we plot the deflection angle $\theta_A$ against the black hole mass $m$ by varying the charge parameter $q$, and impact parameter $b$.  We obtain the following key features of the deflection angle:

 %%%%%%%%%%%%%%%%%%%%%%%%%%%%%%%%%%%%%%%%%%%%%%%%%%%%%%%%%%%%%%%%%%
\subsection{ Deflection angle $\theta_A$ against the impact parameter $b$}
%%%%%%%%%%%%%%%%%%%%%%%%%%%%%%%%%%%%%%%%%%%%%%%%%%%%%%%%%%%%%%%%%%
The behaviour of deflection angle $\theta_A$ against the impact parameter $b$ is depicted in Fig. \ref{fig1} for selecting values of $q$ and $m$.

{\bf Fig. \ref{fig1} (Left)} shows that the deflection angle decreases for increasing impact parameter $b$ in which the black hole mass $m = 2$ and the charge parameter is $q = \{0, 0.4, 0.8, 0.11\} < 0.6 m$. Here, the deflection angle also decreases for increasing values of $q$, i.e. the presence of the charge in the black hole reduced the deflection angle in comparison to the Schwarzschild black hole.

{\bf Fig. \ref{fig1} (Right)} indicates that the deflection angle decreases for increasing impact parameter $b$ in which the black hole mass $m = \{1, 1.5, 2, 2.5\}$, and charge parameter $q = 0.5 < 0.6m$. In addition, the deflection angle increases for increasing values of $m$.

%%%%%%%%%%%%%%%%%%%%%%%%%%%%%%%%%%%%%%%%%%%%%%%%%%%%%%%%%%%%%%%%%%
\subsection{ Deflection angle $\theta_A$ against the charge parameter $q$}
%%%%%%%%%%%%%%%%%%%%%%%%%%%%%%%%%%%%%%%%%%%%%%%%%%%%%%%%%%%%%%%%%%
The behaviour of deflection angle $\theta_A$ against the charge parameter $q$ is displayed in Fig. \ref{fig2} for selecting values of $b$ and $m$.

{\bf Fig. \ref{fig2} (Left)} demonstrates that the deflection angle decreases for increasing charge parameter $q$ in which the black hole mass $m = 2$ and the impact parameter parameter is $b = \{5, 7, 9, 11\}$. Here, the increasing impact parameter $b$ also generates the decreasing deflection angle.

{\bf Fig. \ref{fig2} (Right)} shows that the deflection angle decreases for increasing charge parameter $q$ with the black hole mass $m = \{2, 2.5, 3, 3.5\}$, and impact parameter $b = 6$. Moreover, the deflection angle increases for increasing $m$.

%%%%%%%%%%%%%%%%%%%%%%%%%%%%%%%%%%%%%%%%%%%%%%%%%%%%%%%%%%%%%%%%%%%%%
\begin{figure}[!htbp]
\begin{center}
\begin{tabular}{rl}
\includegraphics[width=8cm]{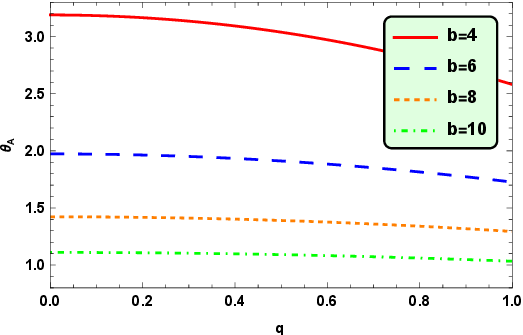}
\end{tabular}
\begin{tabular}{rl}
\includegraphics[width=8cm]{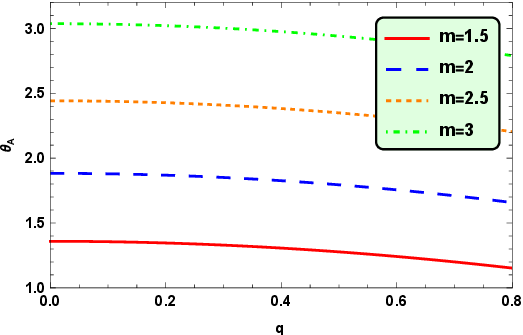}
\end{tabular}
\end{center}
\caption{Deflection angle $\theta_A$ against the charged parameter $q$ corresponding to $m = 2.5$ (Left) and $b = 5$ (Right).}\label{fig5}
\end{figure}
%%%%%%%%%%%%%%%%%%%%%%%%%%%%%%%%%%%%%%%%%%%%%%%%%%%%%%%%%%

%%%%%%%%%%%%%%%%%%%%%%%%%%%%%%%%%%%%%%%%%%%%%%%%%%%%%%%%%%%%%%%%%%
\subsection{ Deflection angle $\theta_A$ against the black hole mass $m$}
%%%%%%%%%%%%%%%%%%%%%%%%%%%%%%%%%%%%%%%%%%%%%%%%%%%%%%%%%%%%%%%%%%
The deflection angle $\theta_A$ against the impact parameter $b$ is shown in Fig. \ref{fig3} for some particular values of $q$ and $b$.

{\bf Fig. \ref{fig3} (Left)} ensure that the deflection angle increases for increasing black hole mass $m$ associated with impact parameter $b = 5$ and the charge parameter is $q = \{0, 0.4, 0.8, 0.11\}$. Also, for the increasing values of $q$, here, the deflection angle decreases .

{\bf Fig. \ref{fig3} (Right)} shows that the deflection angle increases for increasing black hole mass $m$ in which the impact parameter $b = \{6, 8, 10, 12\}$, and charge parameter $q = 0.6$. In this profile, the deflection angle decreases for increasing values of $b$.

%%%%%%%%%%%%%%%%%%%%%%%%%%%%%%%%%%%%%%%%%%%%%%%%%%%%%%%%%%%%%%%%%%
\section{DEflection angle in plasma medium}\label{secVI}
%%%%%%%%%%%%%%%%%%%%%%%%%%%%%%%%%%%%%%%%%%%%%%%%%%%%%%%%%%%%%%%%%%
In this section, we continue to study the influences of plasma medium on the weak deflection angle of light in the gravitational field of the present black hole solution. Regarding this, we take into account  the refractive index for plasma medium, defined as \cite{gc18}
\begin{eqnarray}
    n(r) = \sqrt{1-\frac{\omega_e^2}{\omega_\infty^2}\left[1-\frac{2 m r^2}{\left(q^2+r^2\right)^{3/2}}+\frac{q^2 r^2}{\left(q^2+r^2\right)^2}\right]}
\end{eqnarray}
where $\omega_e$ and $\omega_\infty$ represent the electron and photon plasma frequencies, respectively.

In order to study the effect of plasma medium on the bending angle following Gibbons and Werner's method, we consider the optical geometry of the present black hole (\ref{bh}) with the effect of the plasma medium, written as 
\begin{equation}
dt^2=n(r)^2\left[1-\frac{2 m r^2}{\left(q^2+r^2\right)^{3/2}}+\frac{q^2 r^2}{\left(q^2+r^2\right)^2}\right]^{-2}dr^2+r^2n(r)^2\left[1-\frac{2 m r^2}{\left(q^2+r^2\right)^{3/2}}+\frac{q^2 r^2}{\left(q^2+r^2\right)^2}\right]^{-1} d\phi^2.\label{om}
\end{equation}

The Gaussian optical curvature $\mathcal{G}^{PM}$ for the above optical geometry (\ref{om}) is obtained in the  following form
\begin{eqnarray}
   \mathcal{G}^{PM} &=& -\frac{2 m}{r^3}.+\left(3 m^2+3 q^2-\frac{\omega_e^2 m^2}{\omega_\infty^2}+\frac{\omega_e^2 q^2}{\omega_\infty^2}\right)\frac{1}{r^4}+\left(\frac{15 \omega_e^2 m q^2}{\omega_\infty^2}-\frac{4 \omega_e^4 m^3}{\omega_\infty^4}-\frac{5 \omega_e^4 m q^2}{\omega_\infty^4}-8 m q^2\right)\frac{1}{r^5}+\bigg(\frac{4 \omega_e^4 m^4}{\omega_\infty^4}\nonumber
   \\
   &&+\frac{53 \omega_e^4 m^2 q^2}{\omega_\infty^4}-\frac{24 \omega_e^2 m^2 q^2}{\omega_\infty^2}-\frac{12 \omega_e^2 q^4}{\omega_\infty^2}-3 m^2 q^2+12 q^4\bigg)\frac{1}{r^6}+\bigg(\frac{24 \omega_e^2 m^3 q^2}{\omega_\infty^2}-\frac{110 \omega_e^4 m^3 q^2}{\omega_\infty^4}-\frac{199 \omega_e^4 m q^4}{2 \omega_\infty^4}\nonumber
   \\
   &&+\frac{203 \omega_e^2 m q^4}{4 \omega_\infty^2}+\frac{111 m q^4}{4}\bigg)\frac{1}{r^7}+O\left(\frac{1}{r^8}\right).
\end{eqnarray}

%%%%%%%%%%%%%%%%%%%%%%%%%%%%%%%%%%%%%%%%%%%%%%%%%%%%%%%%%%%%%%%%%%%%%
\begin{figure}[!htbp]
\begin{center}
\begin{tabular}{rl}
\includegraphics[width=8cm]{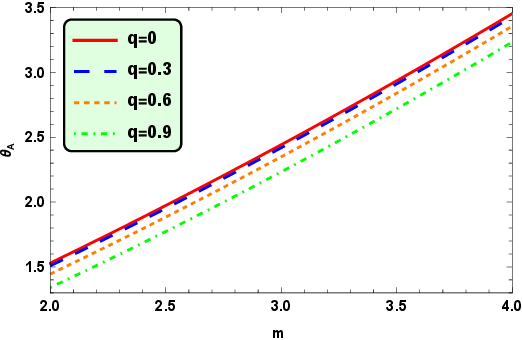}
\end{tabular}
\begin{tabular}{rl}
\includegraphics[width=8cm]{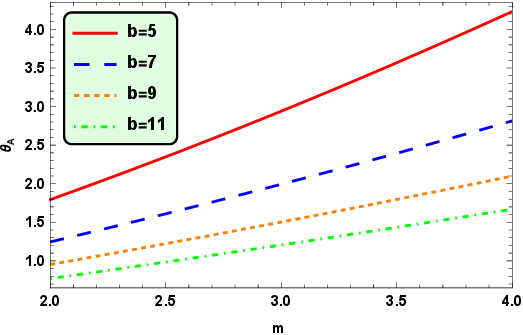}
\end{tabular}
\end{center}
\caption{Deflection angle $\theta_A$ against the mass $m$ corresponding to $b = 6$ (Left) and $q = 0.5$ (Right).}\label{fig6}
\end{figure}
%%%%%%%%%%%%%%%%%%%%%%%%%%%%%%%%%%%%%%%%%%%%%%%%%%%%%%%%%%

Now, we take the surface element $d\mathcal{A}$  as
\begin{eqnarray}
    d\mathcal{A} = \sqrt{g^{optPM}}dr d\phi \simeq r dr d\phi.
\end{eqnarray}

Substituting the above results into the formula for the deflection angle (\ref{da}), we derive the deflection angles in the following forms
\begin{eqnarray}
\theta_A^{PM} &\simeq& \frac{4 m}{b} -\frac{3 \pi  q^2}{4 b^2}-\frac{16 m q^2}{3 b^3}+\frac{9 \pi  q^2 \left(11 m^2+6 q^2\right)}{32 b^4}+\frac{12 m q^4}{25 b^5}+\bigg(\frac{9 \pi  m^2 q^2}{4 b^4 \omega_\infty^2}-\frac{128 m^3 q^2}{25 b^5 \omega_\infty^2}-\frac{812 m q^4}{75 b^5 \omega_\infty^2}+\frac{9 \pi  q^4}{8 b^4 \omega_\infty^2}\nonumber
\\
&&-\frac{20 m q^2}{3 b^3 \omega_\infty^2}+\frac{\pi  m^2}{4 b^2 \omega_\infty^2}-\frac{\pi  q^2}{4 b^2 \omega_\infty^2}\bigg)\omega_e^2.\label{DA1}
\end{eqnarray}

One can see from the above result that the electron and photon plasma frequencies influence the weak deflection angle of light. Moreover, the above deflection angle becomes the deflection angle for a non-plasma medium (\ref{DA}) corresponding to $\omega_e = 0$, i.e. absence of plasma medium. Here, we also examine the behaviors of the deflection angle with the plasma medium effect graphically against the impact parameter, charge parameter and black hole mass. 

%%%%%%%%%%%%%%%%%%%%%%%%%%%%%%%%%%%%%%%%%%%%%%%%%%%%%%%%%%%%%%%%%%
\section{GRAPHICAL ANALYSIS of deflection angle FOR PLASMA MEDIUM}\label{secVII}
%%%%%%%%%%%%%%%%%%%%%%%%%%%%%%%%%%%%%%%%%%%%%%%%%%%%%%%%%%%%%%%%%%
In this section, we explore the deflection angle in the plasma medium graphically to overview the impact of the plasma medium, impact parameter $b$, charge parameter $q$,  and the black hole mass $m$. In this regard, we take into account $\omega_e/\omega_\infty = 1.5$, the variate impact parameter $b$, charge parameter $q$,  and the black hole mass $m$ in order to check the impact of these parameters. The following pivotal outcomes concerning the deflection angle are acquired:

%%%%%%%%%%%%%%%%%%%%%%%%%%%%%%%%%%%%%%%%%%%%%%%%%%%%%%%%%%%%%%%%%%
\subsection{ Deflection angle $\theta_A$ against impact parameter $b$}
%%%%%%%%%%%%%%%%%%%%%%%%%%%%%%%%%%%%%%%%%%%%%%%%%%%%%%%%%%%%%%%%%%
The relationship between the deflection angle $\theta_A$ and the impact parameter $b$ is illustrated in Fig. \ref{fig4}, considering specific values of $q$ and $m$.

In {\bf Fig. \ref{fig4} (Left)}, it is observed that as the impact parameter $b$ increases, the deflection angle decreases. These results are depicted for a black hole mass of $m = 2$ and charge parameters $q = \{0, 0.3, 0.6, 0.9\}$. Additionally, the deflection angle decreases with increasing values of $q$. Moreover, one can see that the deflection angle has increased than the vacuum case due to the effect of the plasma medium. 

In {\bf Fig. \ref{fig4} (Right)}, it is shown that the deflection angle decreases as the impact parameter $b$ increases, with the black hole mass varying as $m = \{ 1.5, 2, 2.5, 3 \}$ and a fixed charge parameter $q = 0.5$. Additionally, the deflection angle increases with increasing values of $m$. Here, the plasma medium increases the deflection angle than the vacuum case.

%%%%%%%%%%%%%%%%%%%%%%%%%%%%%%%%%%%%%%%%%%%%%%%%%%%%
\begin{table*}[thth]
\caption{ Numerical values of inter and outer horizons $r_{\pm}$, radius of photon sphere $r_{ps\pm}$ corresponding to $r_{\pm}$, and shadow radius $r_{sh\pm}$ for corresponding to $r_{ps\pm}$.} \label{tab1}
\label{tab2}       % Give a unique label
\centering
\begin{tabular}{|c|c|c|c|c|c|c|c|c|c|c|c|c|c|c|c|c|}
\hline
 \multicolumn{7}{|c|}{ m = 1} & \multicolumn{7}{|c|}{q = 0.3}\\
\hline
q & $r_-$ &   $r_+$       &  $r_{ps-}$   &  $r_{ps+}$ &   $r_{sh-}$ & $r_{sh+}$ & m & $r_-$ &   $r_+$       &  $r_{ps-}$   &  $r_{ps+}$ &   $r_{sh-}$ & $r_{sh+}$    \\
\hline
0    & 2  &  2  &  3   & 3       & 5.19615   & 5.19615        & 1    & 0.146701  &  1.88049  &  0.239609   & 2.85798  & complex  & 5.03345\\
\hline
0.2  & 0.0729727  &  1.94871  &  0.130751  &  2.93869 & complex & 5.12559   & 1.25     & 0.124068  &  2.40658  &  0.213405   & 3.63854  & complex  & 6.36711\\
\hline
0.4 & 0.252294  &  1.77517  &  0.38287   &  2.73569 & complex & 4.89571    & 1.5    & 0.109459  &  2.92307  & 0.196127  & 4.40804  & complex  & 7.68839\\
\hline
0.55 & 0.521539  &  1.49732  & 0.70230   & 2.43511  & complex   & 4.56997  & 1.75     & 0.0990314  &  3.43451  & 0.183583   & 5.17164  & complex  & 9.00299\\
\hline
\end{tabular}\label{table1}
\end{table*}
%%%%%%%%%%%%%%%%%%%%%%%%%%%%%%%%%%%%%%%%%%%%%%%%%%%%%%%%%%%%%%%%%%%%%%%%%%

%%%%%%%%%%%%%%%%%%%%%%%%%%%%%%%%%%%%%%%%%%%%%%%%%%%%%%%%%%%%%%%%%%
\subsection{ Deflection angle $\theta_A$ against charge parameter $q$}
%%%%%%%%%%%%%%%%%%%%%%%%%%%%%%%%%%%%%%%%%%%%%%%%%%%%%%%%%%%%%%%%%%
The nature of deflection angle $\theta_A$ against the charge parameter $q$ is illustrated in Fig. \ref{fig5} for selecting values of $b$ and $m$.

In {\bf Fig. \ref{fig5} (Left)}, it is shown that the deflection angle decreases with increasing charge parameter $q$, with the black hole mass fixed at $m = 2.5$ and impact parameters $b = \{4, 6, 8, 10\}$. Additionally, increasing the impact parameter $b$ also leads to a decrease in the deflection angle. The presence of a plasma medium also increases the deflection angle compared to the vacuum scenario.

In {\bf Fig. \ref{fig5} (Right)}, it is observed that the deflection angle decreases with increasing charge parameter $q$, for fixed values of black hole mass $m = {1.5, 2, 2.5, 3}$, and impact parameter $b = 5$. Furthermore, the deflection angle increases with increasing values of $m$. Here, the deflection angle is also higher in the presence of a plasma medium compared to the vacuum scenario.

%%%%%%%%%%%%%%%%%%%%%%%%%%%%%%%%%%%%%%%%%%%%%%%%%%%%%%%%%%%%%%%%%%
\subsection{ Deflection angle $\theta_A$ against black hole mass $m$}
%%%%%%%%%%%%%%%%%%%%%%%%%%%%%%%%%%%%%%%%%%%%%%%%%%%%%%%%%%%%%%%%%%
The deflection angle $\theta_A$ against the black hole mass $m$ is demonstrated in Fig. \ref{fig6} corresponding to some specific values of $q$ and $b$.

In {\bf Fig. \ref{fig6} (Left)}, it is evident that the deflection angle increases with increasing black hole mass $m$, considering a fixed impact parameter $b = 6$ and varying charge parameter $q = \{0, 0.3, 0.6, 0.9\}$. Additionally, as the value of $q$ increases, the deflection angle decreases.

In {\bf Fig. \ref{fig6} (Right)}, it is depicted that the deflection angle increases with increasing black hole mass $m$, with the impact parameter set to $b = \{5, 7, 9, 11\}$ and a fixed charge parameter $q = 0.5$. Additionally, within this setup, the deflection angle decreases as the impact parameter $b$ increases.

In both these scenarios,  the presence of a plasma medium results in a higher deflection angle compared to the vacuum case.

%%%%%%%%%%%%%%%%%%%%%%%%%%%%%%%%%%%%%%%%%%%%%%%%%%%%%%%%%%%%%%%%%%
\section{Time Delay}\label{secVIII}
%%%%%%%%%%%%%%%%%%%%%%%%%%%%%%%%%%%%%%%%%%%%%%%%%%%%%%%%%%%%%%%%%%
In this section, we study the time delay in the gravitational field of the black hole (\ref{bh}). The time delay refers to the difference in time between two photons released simultaneously from the source, travelling along different paths toward the viewer. Suppose light propagates through a static spherically symmetric spacetime, defined as
\begin{eqnarray}
    ds^2 = -A(r)dt^2 +B(r)dr^2+C(r)(d\theta^2+\sin^2\theta d\phi^2),
\end{eqnarray}

 Then the time delay of the light can be defined as \cite{weinberg1} 
\begin{eqnarray}
\Delta T = 2\int_{r_{c}}^{r_{c}}\left[\frac{1}{\sqrt{\left[\frac{A(r)}{B(r)}-\frac{A^2(r)}{B(r)C(r)}\frac{C(r_c)}{A(r_c)}\right]}}-\frac{1}{\sqrt{\left[1-\frac{r_c^2}{r^2}\right]}}\right]dr
+ 2\int_{r_{0}}^{r_{s}}\left[\frac{1}{\sqrt{\left[\frac{A(r)}{B(r)}-\frac{A^2(r)}{B(r)C(r)}\frac{C(r_c)}{A(r_c)}\right]}}-\frac{1}{\sqrt{\left[1-\frac{r_c^2}{r^2}\right]}}\right]dr.\label{TD}
\end{eqnarray}
Here, $r_v$ and $r_s$ denote the distances of the viewer and the source from the considered astrophysical object, respectively while $r_c$ denotes the closest approach to that object.

Now, let us suppose that the light signal travels through the gravitational field of the black hole (\ref{bh}), then we have 
\begin{eqnarray}
\mathcal{I} &=&\left[\frac{A(r)}{B(r)}-\frac{A^2(r)}{B(r)C(r)}\frac{C(r_c)}{A(r_c)}\right]^{-\frac{1}{2}}=\left[1-\frac{2 m r^2}{\left(q^2+r^2\right)^{3/2}}+\frac{q^2 r^2}{\left(q^2+r^2\right)^2}\right]^{-1}\left[1-\frac{r_c^2}{r^2}\frac{1-\frac{2 m r^2}{\left(q^2+r^2\right)^{3/2}}+\frac{q^2 r^2}{\left(q^2+r^2\right)^2}}{1-\frac{2 m r_c^2}{\left(q^2+r_c^2\right)^{3/2}}+\frac{q^2 r_c^2}{\left(q^2+r_c^2\right)^2}}\right]^{-\frac{1}{2}}
\\
 &\approx& \left[1-\frac{r _{c} ^{2}}{r^{2}}\right]^{-\frac{1}{2}}\left[1+\frac{2 m r^2}{\left(q^2+r^2\right)^{3/2}}-\frac{q^2 r^2}{\left(q^2+r^2\right)^2}-\frac{m r_c^2 r^2}{\left(q^2+r^2\right)^{3/2}}+\frac{r_c^2 q^2}{2 \left(q^2+r^2\right)^2}\right].
\end{eqnarray}

After imposing the above approximate values of the $\mathcal{I}$ in Eq. (\ref{TD})  the time delay reads as
\begin{eqnarray}
\Delta T &=&  2\int_{r_{c}}^{r_{e}}\left[1-\frac{r _{c} ^{2}}{r^{2}}\right]^{-\frac{1}{2}}\bigg[\frac{2 m r^2}{\left(q^2+r^2\right)^{3/2}}-\frac{q^2 r^2}{\left(q^2+r^2\right)^2}-\frac{mr_c^2 r^2}{\left(q^2+r^2\right)^{3/2}}+\frac{r_c^2 q^2}{2 \left(q^2+r^2\right)^2}\bigg]dr
\nonumber
\\
&&+ 2\int_{r_{c}}^{r_{s}}\left[1-\frac{r _{c} ^{2}}{r^{2}}\right]^{-\frac{1}{2}}\bigg[\frac{2 m r^2}{\left(q^2+r^2\right)^{3/2}}-\frac{q^2 r^2}{\left(q^2+r^2\right)^2}-\frac{mr_c^2 r^2}{\left(q^2+r^2\right)^{3/2}}+\frac{r_c^2 q^2}{2 \left(q^2+r^2\right)^2}\bigg]dr
\end{eqnarray}
Finally,  we obtain the time delay in the following form
\begin{eqnarray}
\Delta T&=&\frac{q^2 \left(3 r_c^2+2 q^2\right)}{\left(r_c^2+q^2\right)^{3/2}}T_1-8 m T_2-\frac{\left(r_c^2+2 q^2\right)}{2(r_c^2+q^2)}T_3 + 4m \log \left(r_c^2+q^2\right)+\frac{ q^2 \left(3 r_c^2+2 q^2\right)}{\left(r_c^2+q^2\right)^{3/2}}-\frac{\pi  q^2 \left(3 r_c^2+2 q^2\right)}{2\left(r_c^2+q^2\right)^{3/2}}.
\end{eqnarray}
where
\begin{eqnarray}
    T_1 &=& \tan ^{-1}\left(\frac{\sqrt{q^2+r_v^2}-\sqrt{r_v^2-r_c^2}}{\sqrt{r_c^2+q^2}}\right)+\tan ^{-1}\left(\frac{\sqrt{q^2+r_s^2}-\sqrt{r_s^2-r_c^2}}{\sqrt{r_c^2+q^2}}\right),\nonumber
    \\
 T_2 &=& \log \left(\sqrt{r_v^2-r_c^2}+\sqrt{q^2+r_v^2}\right)+\log \left(\sqrt{r_s^2-r_c^2}+\sqrt{q^2+r_s^2}\right), \nonumber
 \\
 T_3 &=& \frac{\sqrt{r_v^2-r_c^2} \left(q^2-4 m \sqrt{q^2+r_v^2}\right)}{q^2+r_v^2}+\frac{\sqrt{r_s^2-r_c^2} \left(q^2-4 m \sqrt{q^2+r_s^2}\right)}{q^2+r_s^2}.
\end{eqnarray}

It is important to note that the time delay becomes zero as desired for $m = 0$ and $q = 0$, i.e., the absence of the black hole.

%%%%%%%%%%%%%%%%%%%%%%%%%%%%%%%%%%%%%%%%%%%%%%%%%%%%%%%%%%%%%%%%%%
\section{Shadow cast}\label{secIX}
%%%%%%%%%%%%%%%%%%%%%%%%%%%%%%%%%%%%%%%%%%%%%%%%%%%%%%%%%%%%%%%%%%
In general, a photon emitted from a light source tends to be deflected as it nears a black hole due to the effect of gravitational lensing. Indeed,  some photons can eventually reach a distant observer after being deflected by the black hole, while others may directly plunge into the black hole and form the shadow of the black hole. To study shadow formation, we consider the motion of test particles around the present black hole (\ref{bh}) and the Hamilton-Jacobi equation, defined as \cite{bc68a}
\begin{eqnarray}
    \frac{\partial S}{\partial \sigma} =-\frac{1}{2}g^{\alpha\beta}\frac{\partial S}{\partial x^\alpha} \frac{\partial S}{\partial x^\beta}
\end{eqnarray}
 where $S$ represents the Jacobi action. For the photon, the separable solution of the Jacobi action reads as
\begin{eqnarray}
    S = -Et+l\phi+S_r(r)+S_\theta(\theta),
\end{eqnarray}
where $E$ and $l$ are the two Killing vectors, can be given for the metric (\ref{bh}) as
\begin{eqnarray}
    E &=&\frac{dL}{dt} = - \left(1-\frac{2 m r^2}{\left(q^2+r^2\right)^{3/2}}+\frac{q^2 r^2}{\left(q^2+r^2\right)^2}\right)\dot{t},~~~~~~~ l= \frac{dL}{d\phi} = r^2 \sin^2\theta \dot{\phi}.
\end{eqnarray}

%%%%%%%%%%%%%%%%%%%%%%%%%%%%%%%%%%%%%%%%%%%%%%%%%%%%%%%%%%%%%%%%%%%%%
\begin{figure}[!htbp]
\begin{center}
\begin{tabular}{rl}
\includegraphics[width=8cm]{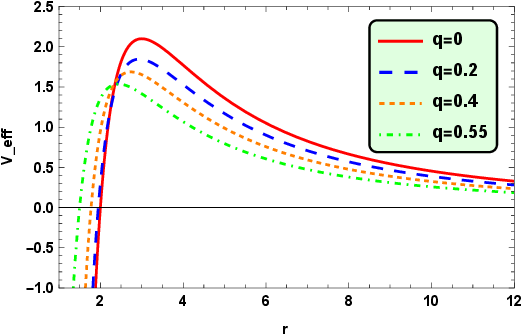}
\end{tabular}
\begin{tabular}{rl}
\includegraphics[width=8cm]{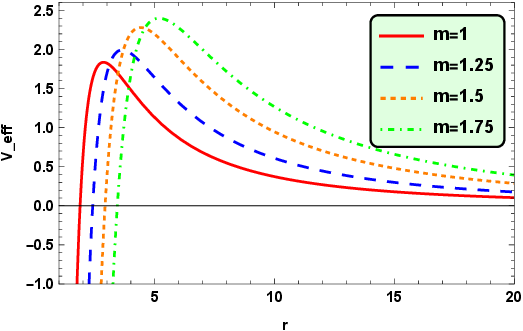}
\end{tabular}
\end{center}
\caption{Effective potential again the radial coordinate $r$ corresponding to $m = 1$ (Left) and $q = 0.3$ (Right).}\label{fig7a}
\end{figure}
%%%%%%%%%%%%%%%%%%%%%%%%%%%%%%%%%%%%%%%%%%%%%%%%%%%%%%%%%%

Therefore, the geodesic equations can be written as
\begin{eqnarray}
    \frac{dt}{d\sigma} &=& \frac{E}{\left(1-\frac{2 m r^2}{\left(q^2+r^2\right)^{3/2}}+\frac{q^2 r^2}{\left(q^2+r^2\right)^2}\right)},
    \\
    \frac{d\phi}{d\sigma} &=& -\frac{l}{r^2 \sin^2\theta},
    \\
    \frac{dr}{d\sigma}&=&\pm \frac{\sqrt{\mathcal{R}(r)}}{r^2},
    \\
    \frac{d\theta}{d\sigma}&=&\pm \frac{\sqrt{\Theta(\theta)}}{r^2},
\end{eqnarray}

where
\begin{eqnarray}
    \mathcal{R}(r) = r^4E^2-(\mathcal{K}+l^2)r^2 \left(1-\frac{2 m r^2}{\left(q^2+r^2\right)^{3/2}}+\frac{q^2 r^2}{\left(q^2+r^2\right)^2}\right),~~~\Theta(\theta) = \mathcal{K}-l^2 \cot\theta,
\end{eqnarray}
with $\mathcal{K}$ as the Carter separation constant \cite{bc68}. 

Here, we are interested in the spherical photon orbit which is the spherical light path constrained on a sphere of fixed radius $r$ with $\dot{r} = 0$ and $\ddot{r} = 0$. The fixed radius $r = r_{ps}$ of the spherical photon orbit is known as the radius of the photon sphere, and it indicates the position of the apparent image of the photon rings. For a distant observer, photons approaching the described black hole (\ref{bh}) near its equatorial plane and the unstable circular orbits follow
\begin{eqnarray}
\mathcal{R}_{eff}(r)|_{r=r_{ps}} = 0,~~~ \mathcal{R}'_{eff}(r)|_{r=r_{ps}} = 0.
\end{eqnarray}

Now, we introduced two dimensionless impact parameters $\eta$ and $\xi$ depending on the values of critical parameters whether the photon is captured, scattered to infinity, or bound to orbits, defined as
\begin{eqnarray}
    \eta = \frac{\mathcal{K}}{E^2},~~~~\xi = \frac{l}{E}.
\end{eqnarray} 

Therefore, the solution $\mathcal{R}(r)$ can be formulated in terms of these dimensionless impact parameters as
\begin{eqnarray}
    \mathcal{R}(r) = r^4E^2-r^2 E^2(\xi^2+\eta)\left(1-\frac{2 m r^2}{\left(q^2+r^2\right)^{3/2}}+\frac{q^2 r^2}{\left(q^2+r^2\right)^2}\right).
\end{eqnarray}

%%%%%%%%%%%%%%%%%%%%%%%%%%%%%%%%%%%%%%%%%%%%%%%%%%%%%%%%%%%%%%%%%%%%%
\begin{figure}[!htbp]
\begin{center}
\begin{tabular}{rl}
\includegraphics[width=6cm]{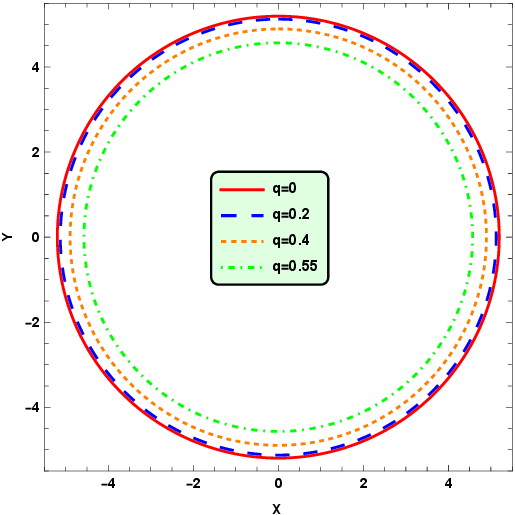}
\end{tabular}
\begin{tabular}{rl}
\includegraphics[width=6cm]{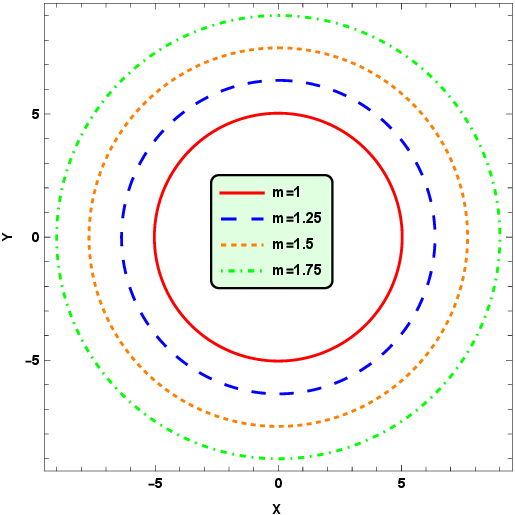}
\end{tabular}
\end{center}
\caption{Shadows of the black hole corresponding to $m = 1$ (Left) and $q = 0.3$ (Right).}\label{fig7}
\end{figure}
%%%%%%%%%%%%%%%%%%%%%%%%%%%%%%%%%%%%%%%%%%%%%%%%%%%%%%%%%%

To analyse the path of photons, it is advantageous to express the radial geodesics with the effective potential $V_{eff}(r)$ as
\begin{eqnarray}
    \frac{dr}{d\sigma} + V_{eff}(r)= 0,
\end{eqnarray}
where
\begin{eqnarray}
    V_{eff}(r)= -\frac{1}{E^2 r^4} \mathcal{R}(r) =-\frac{1}{r^2} \left[r^2-(\xi^2+\eta)\left(1-\frac{2 m r^2}{\left(q^2+r^2\right)^{3/2}}+\frac{q^2 r^2}{\left(q^2+r^2\right)^2}\right)\right].
\end{eqnarray} 

 The unstable circular orbits with maximum effective potential $V_{eff}$ follow the conditions
\begin{eqnarray}
    V_{eff}(r)|_{r=r_{ps}} = 0,~~~ V'_{eff}(r)|_{r=r_{ps}} = 0,\label{v1}
\end{eqnarray}

We have analyzed the radial dependence of the effective potential corresponding to black hole mass $m = 1$ and charge parameter $q=\{0, 0.2, 0.4, 0.55\}$ in Fig. \ref{fig7a} (Left) and black hole mass $m = \{1, 1.25, 1.5, 1.75\}$ and charge parameter $q=0.3$ in Fig. \ref{fig7a} (Right). Here, we can see that the maximum value of the effective potential corresponding to photon orbit is decreased and shifted towards the black hole with the increase of the value of charge parameter $q$ (See Fig. \ref{fig7a} (Left)) while the maximum value of the effective potential has opposite behaviour for increasing the black mass $m$ (See Fig. \ref{fig7a} (Right)). The maximum effective potential admits an unstable circular orbit, thus, photons can exit their circular orbits due to disturbances caused by external gravitational forces or through interactions with other particles, constructing a photon sphere that is observable as a black hole shadow by the observer.   

Now, the first condition of the above Eq. (\ref{v1}) yields the following result
\begin{eqnarray}
\xi^2+\eta = \frac{r_{ps}^2}{1-\frac{2 m r_{ps}^2}{\left(q^2+r_{ps}^2\right)^{3/2}}+\frac{q^2 r_{ps}^2}{\left(q^2+r_{ps}^2\right)^2}}.
\end{eqnarray}

 Also, both the conditions of Eq. (\ref{v1}) simultaneously lead to the following result  for our described black hole (\ref{bh}) 
\begin{eqnarray}
    r_{ps}^{12}+\left(10 q^2-9 m^2\right)r_{ps}^{10} +\left(31 q^4-9 m^2 q^2\right)r_{ps}^8 +32 q^6 r_{ps}^6+19 q^8 r_{ps}^4+6 q^{10} r_{ps}^2+q^{12}= 0.\label{ps}
\end{eqnarray}
The above Eq. (\ref{ps}) has twelve roots, and only two of these are real positive, given in the table-\ref{tab1} by solving numerically.

Now, the celestial coordinates $X$ and $Y$ can be used to describe the shadow of the black hole, defined as \cite{sv04}
\begin{eqnarray}
    X &=& \lim_{r_0\rightarrow \infty}\left(-r_0\sin\theta_0\left[\frac{d\phi}{dr}\right]_{r_0,\theta_0}\right),\label{x}~~~~
    Y = \lim_{r_0\rightarrow \infty}\left(r_0\left[\frac{d\theta}{dr}\right]_{r_0,\theta_0}\right),\label{y}
\end{eqnarray}
where $(r_0, \theta_0)$ are the position coordinates of the observer. For the null geodesic, the above coordinate reads as 
\begin{eqnarray}
    X &=& -\frac{\xi}{\sin\theta},\label{x1}~~~~~ Y = \pm \sqrt{\eta -\xi^2 \cot^2\theta},\label{y1}
\end{eqnarray}

Suppose that the observer is on the equatorial plane $(\theta = \pi/2)$, the celestial coordinates (\ref{x}) and (\ref{y}) take the values
\begin{eqnarray}
    X &=& -\xi, ~~~~~ Y = \pm \sqrt{\eta}.
\end{eqnarray}

Therefore, the radius of the shadow $r_{sh}$ for our described black hole (\ref{bh}) can be expressed as
\begin{eqnarray}
    r_{sh} = \sqrt{X^2 + Y^2} = \sqrt{\xi^2 +\eta}= \frac{r_{ps}}{\sqrt{1-\frac{2 m r_{ps}^2}{\left(q^2+r_{ps}^2\right)^{3/2}}+\frac{q^2 r_{ps}^2}{\left(q^2+r_{ps}^2\right)^2}}} .
\end{eqnarray}

It is noted that the concept of a black hole shadow pertains to the area outside the outer event horizon, where light can either escape or be captured, forming a discernible shadow visible from a distance. In contrast, the inner event horizon lacks a corresponding observable shadow radius. This distinction arises because the inner event horizon resides within the intricate interior of the black hole, which is inaccessible to direct observation from external vantage points due to the complexities of spacetime structure. For the present black hole, the shadow radii are given in Table-\ref{tab1}. Interestingly, no real shadow radius exists for the inner event horizon.  The shadows of our described black hole are displayed in Fig. \ref{fig7}  for selected values charge parameter $q$ and black hole mass $m$. One can see that the size of the black hole shadow is decreased with increasing parameter charge parameter $q$ (See Fig. \ref{fig7} (Left)) and increased for increasing mass $m$ (See Fig. \ref{fig7} (Right)).

%%%%%%%%%%%%%%%%%%%%%%%%%%%%%%%%%%%%%%%%%%%%%%%%%%%%%%%%%%%%%%%%%%%%%
\begin{figure}[!htbp]
\begin{center}
\begin{tabular}{rl}
\includegraphics[width=8cm]{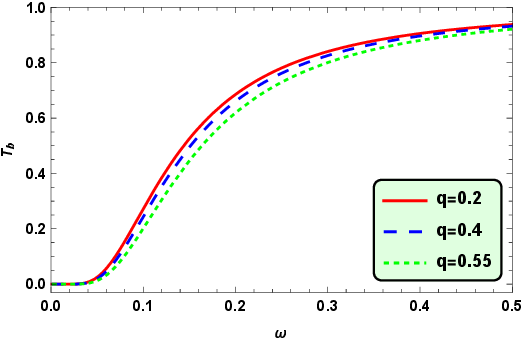}
\end{tabular}
\begin{tabular}{rl}
\includegraphics[width=8cm]{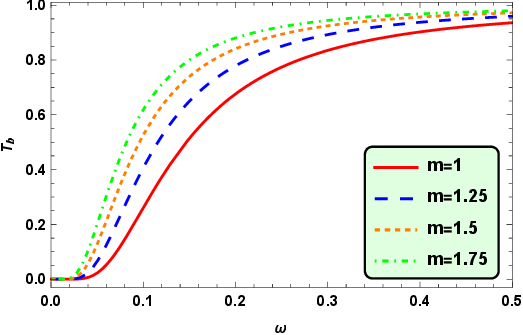}
\end{tabular}
\end{center}
\caption{Greybody factor $\mathcal{T}_b$ against the the quasinormal mode frequency $\omega$ corresponding to $m = 1$ (Left) and $q = 0.3$ (Right).}\label{fig8}
\end{figure}
%%%%%%%%%%%%%%%%%%%%%%%%%%%%%%%%%%%%%%%%%%%%%%%%%%%%%%%%%%

%%%%%%%%%%%%%%%%%%%%%%%%%%%%%%%%%%%%%%%%%%%%%%%%%%%%%%%%%%%%%%%%%%
\section{Bound OF GREYBODY factor}\label{secX}
%%%%%%%%%%%%%%%%%%%%%%%%%%%%%%%%%%%%%%%%%%%%%%%%%%%%%%%%%%%%%%%%%%
In this section, we study the rigorous bound of the greybody factor (GBF) of the present black hole (\ref{bh}). Indeed, the GBF is related to the quantum nature of a black hole that describes its emissivity. The general bounds of the GBF can be expressed as \cite{mv99}
\begin{eqnarray}
    \mathcal{T}_b \geq sech^2\left(\frac{1}{2\omega}\int_{-\infty}^{\infty} \mathcal{V}(r)dr^*\right),\label{T}
\end{eqnarray}
where $\omega$ represents the quasinormal mode frequency, and  $r^*$ stands for the tortoise coordinate. 

 The Regge–Wheeler equation with the angular momentum $\mathcal{L}$ is defined as
\begin{eqnarray}
   \left[\frac{d^2}{dr^*2}+\omega^2-\mathcal{V}(r)\right]\psi = 0,
\end{eqnarray}
where 
\begin{eqnarray}
   dr^* = \frac{1}{g_{rr}}dr,\label{dr}
\end{eqnarray}
and the potential $\mathcal{V}(r)$ for the four-dimensional spacetime is defined as 
\begin{eqnarray}
    \mathcal{V}(r) = \frac{g_{rr}}{r}\partial_r g_{rr}+\mathcal{L}(\mathcal{L}+1)\frac{g_{rr}}{r^2}.\label{v}
\end{eqnarray}

After imposing the above results (\ref{dr}) and (\ref{v}), the lower bound value of GBF (\ref{T}) reads as 
\begin{eqnarray}
    \mathcal{T}_b \geq sech^2\left(\frac{1}{2\omega}\int_{r_+}^{\infty} \left[ \frac{1}{r}\partial_r g_{rr}+\mathcal{L}(\mathcal{L}+1)\frac{1}{r^2}\right]dr\right).\label{T2}
\end{eqnarray}

Therefore, for the desired black hole (\ref{bh}), we obtain the lower bound of GBF in the following form
\begin{eqnarray}
    \mathcal{T} \geq sech^2\left(\frac{1}{2\omega} \left[ \frac{\pi }{4 q}-\frac{2 m}{q^2}+\frac{\mathcal{L}(\mathcal{L}+1)}{ r_+}+\frac{r_+}{2} \left(\frac{4 m \left(2 q^2+r_+^2\right)}{q^2 \left(q^2+r_+^2\right)^{3/2}}-\frac{3 q^2+r_+^2}{\left(q^2+r_+^2\right)^2}\right)-\frac{1}{2 q}\tan ^{-1}\left(\frac{r_+}{q}\right)\right]\right).\label{T2}
\end{eqnarray}

We display the lower bound against the quasinormal mode frequency $\omega$ by considering $\mathcal{L} = 0$, and the specific values of $q = \{0.2, 0.4, 0.55\}$, $m = 1$ in Fig. \ref{fig8} (Left) and  $m = \{1, 1.25, 1.5, 1.75\}$, $q = 0.3$ in Fig. \ref{fig8} (Right) along with the respective outer horizon $r_+$ given in Table-\ref{tab1}. Fig. \ref{fig8} (Left)  shows that the estimated lower bound of the described black hole is decreasing for increasing variation of charge parameter $q$. Consequently, the obtained $\mathcal{T}_b$ has an inverse relation with the charge parameter $q$. Besides, the lower bound of the black hole increases with the increasing values of mass $m$, i.e. $\mathcal{T}_b$, and the mass $m$ has a direct relationship (See Fig. \ref{fig8} (Right)). Therefore, it can be concluded that the higher values of charge parameter $q$  result in the lowering of the GBF $\mathcal{T}_b$ while the higher values of black hole mass $m$  provide the higher values of the GBF $\mathcal{T}_b$. It is worth mentioning that the above process will give the bound of GBF for the Schwarzschild black hole corresponding to $q =0$ \cite{pb08}.

%%%%%%%%%%%%%%%%%%%%%%%%%%%%%%%%%%%%%%%%%%%%%%%%%%%%%%%%%%%%%%%%%%
\section{Results and conclusion}\label{secXI}
%%%%%%%%%%%%%%%%%%%%%%%%%%%%%%%%%%%%%%%%%%%%%%%%%%%%%%%%%%%%%%%%%%

 In recent years, the study of black hole spacetimes has gained remarkable attention, particularly in the context of gravitational lensing across various gravitational theories. Gravitational lensing remains one of the most powerful astrophysical probes for exploring spacetime geometry and testing alternative models of gravity. Motivated by this, our work investigates the weak deflection angles, time delay, shadow, and greybody factors associated with the regular, static, spherically symmetric black holes in the framework of nonlinear electrodynamics. The analysis of weak deflection angles in such regular black holes not only provides deeper theoretical insights into the interplay between regularity and nonlinear electrodynamics but also highlights potential observational signatures that could distinguish regular black holes from their singular counterparts. We have conducted this study in the context of vacuum medium as well as plasma medium to examine the impact of plasma medium on the deflection angle. In this regard, we have considered the propagation of light rays at the equatorial plane to obtain the corresponding optical metrics and Gaussian optical curvatures of the present black holes, both in the vacuum medium and the plasma medium. Accordingly, we have estimated the deflection angles of light $\theta_A$ in the vacuum medium and $\theta_A^{PM}$ in the plasma medium by using the very well-known Gauss-Bonnet method  in the weak field limit, found as 
\begin{eqnarray}
\theta_A &\simeq& \frac{4 m}{b} -\frac{3 \pi  q^2}{4 b^2}-\frac{16 m q^2}{3 b^3}+\frac{9 \pi  q^2 \left(11 m^2+6 q^2\right)}{32 b^4}+\frac{12 m q^4}{25 b^5}.\label{DA}
\\
\theta_A^{PM} &\simeq& \frac{4 m}{b} -\frac{3 \pi  q^2}{4 b^2}-\frac{16 m q^2}{3 b^3}+\frac{9 \pi  q^2 \left(11 m^2+6 q^2\right)}{32 b^4}+\frac{12 m q^4}{25 b^5}+\bigg(\frac{9 \pi  m^2 q^2}{4 b^4 \omega_\infty^2}-\frac{128 m^3 q^2}{25 b^5 \omega_\infty^2}-\frac{812 m q^4}{75 b^5 \omega_\infty^2}+\frac{9 \pi  q^4}{8 b^4 \omega_\infty^2}\nonumber
\\
&&-\frac{20 m q^2}{3 b^3 \omega_\infty^2}+\frac{\pi  m^2}{4 b^2 \omega_\infty^2}-\frac{\pi  q^2}{4 b^2 \omega_\infty^2}\bigg)\omega_e^2.\label{DA1}
\end{eqnarray}

It is worth mentioning that if we choose the electric charge $q = 0$ in the above results, our estimated deflection angles are reduced to the deflection angles $\frac{4 m}{b}$  in the vacuum medium and $\frac{4 m}{b} +\frac{\pi  m^2}{4 b^2 }\frac{\omega_e^2}{\omega_\infty^2}$ in the plasma medium for Schwarzschild black hole up to the first order term. It is also noted that after removing the plasma effect, i.e. $\omega_e = 0$ or $\omega_e/\omega_\infty\rightarrow 0$, the deflection angle in the plasma medium reduces to the result of the vacuum medium, indicating the consistency of the obtained deflection angles. In addition, we have graphically analyzed the effect of the impact parameter $b$, charge parameter $q$, and black hole mass $m$ on both the obtained deflection angles in Figs. \ref{fig1}-\ref{fig6}, and found the following results:  
\begin{itemize}
\item Both deflection angles decrease gradually as the impact parameter $b$ increases,  which implies that light rays passing closer to the black hole (smaller $b$) experience stronger curvature of spacetime, resulting in larger deflection angles, while those with larger impact parameters (farther trajectories from the black hole) are only weakly influenced by the gravitational field, leading to smaller bending angles. A similar dependence of the deflection angle on the impact parameter has also been reported for various black hole spacetimes in Refs. \cite{WJ22, yk23, rk19, yk20, rc20, ao19a, ab22, kj18, wj19}. 

\item  Both deflection angles are gradually decreasing for increasing values of charge parameter $q$.  This nature of deflection angles indicates that the presence of charge modifies the spacetime geometry of the black hole in such a way that the effective gravitational attraction experienced by light rays is reduced. In other words, a higher value of $q$ weakens the curvature of spacetime compared to the purely Schwarzschild case, leading to smaller bending of light trajectories. This behavior reflects the fact that the electromagnetic repulsion introduced by the charge counteracts, to some extent, the gravitational pull, thereby diminishing the overall deflection angle. Notably, a similar behavior of the deflection angle is observed for the Bardeen black hole, as reported in Ref.~\cite{hg16}.

\item  Both deflection angles increase gradually with increasing values of the black hole mass $m$. This is because a larger mass enhances the gravitational field strength, leading to stronger curvature of spacetime. As a result, light rays passing near the black hole experience greater bending compared to those around a lighter black hole. This dependence of the deflection angle on mass $m$ reflects the direct role of mass in governing the intensity of gravitational lensing. This result is consistent with the findings reported in Ref. \cite{wj22}.

\end{itemize}

Moreover, the presence of the plasma medium has increased the deflection angle as compared to the vacuum medium. This effect arises because the plasma behaves as a dispersive medium, altering the refractive index along the photon’s path. Thus, light rays propagating through plasma are influenced not only by the gravitational curvature of spacetime but also by plasma-induced refraction, which adds an extra bending component. Consequently, the deflection angle in the plasma medium for the present study is larger than in vacuum, reflecting the combined action of gravitational and medium-induced effects. This result is consistent with the findings reported in Refs. \cite{AO19, fu21, fa23}. Besides, we have calculated the mathematical expression for the time delay in the background of present black holes, which allows us to analyze the difference in arrival times between two photons emitted simultaneously from a source but traveling along distinct trajectories toward the observer. Interestingly, the time delay as expected becomes zero in the absence of present black holes, i.e. when $m = 0$ and $q = 0$. Also, to study the shadow cast of the black hole, we have estimated the radius of the inner horizon and outer horizon, the radius of the photon sphere, and the shadow radius for some specific values of charge parameter $q$ and black hole mass $m$, given in Table-\ref{tab1}. For the inner and outer horizons, we have obtained two photon-sphere radii, $r_{ps-}$ and $r_{ps+}$, respectively. However, when $q \neq 0$, no physical shadow radius exists for $r_{ps-}$, whereas a real shadow radius $r_{sh+}$ is obtained for $r_{ps+}$. For $m = 1$, and $q = 0$, we have found the event horizon $r_- = r_+ = 2$, photon-sphere radius $r_{ps-} = r_{ps+} = 3$, and the shadow radius $r_{sh-}$ = $r_{sh+}$  = 5.19615, which are in agreement with the results for the Schwarzschild black hole reported in Ref. \cite{jp79}. In Fig. \ref{fig7} (left), we have demonstrated the behavior of the shadow with the variation of $q$, which shows that the shadow region decreases for increasing values of $q$, while the shadow region expands as the value of $m$ increases, clear in Fig. \ref{fig7} (Right). We have also calculated the lower bound result of the GBF $\mathcal{T}_b$ for the described black hole, and it is found that increasing the charge parameter $q$ leads to a decrease in the GBF, whereas increasing the black hole mass $m$ results in higher values of the GBF.  Therefore, the novelty of this study lies in extending the application of the Gauss–Bonnet theorem to determine the weak deflection angles by the regular black holes arising from nonlinear electrodynamics. All the results obtained in this study highlight the role of the mass $m$ and non-singularity of the black holes through the charge parameter $q$. The mass enriches all the findings by strengthening the curvature of spacetime, whereas the non-singularity via the charge parameter $q$ diminishes the findings by diminishing the curvature of the spacetime.

Finally, as a concluding remark, we believe that the results we have obtained will provide insights for future observations that can provide more substantial information on the impact of these parameters on the deflection angles.

\section*{Acknowledgement}
FR would like to thank the authorities of the Inter-University Centre for Astronomy and Astrophysics, Pune, India, for providing research facilities. FR is also thankful to SERB, DST  $\&$   DST FIST programme  (SR/FST/MS-II/2021/101(C))  for financial support. We are deeply grateful to the reviewer(s) for their valuable and constructive suggestions that have enriched this manuscript.

\end{document}